# Reversal of charge transfer doping on the negative electronic compressibility surface of MoS$_2$

Liam Watson[1,2], Iolanda Di Bernardo[1,2,3,4], James Blyth[1,2], Benjamin Lowe[1,2], Thi-Hai-Yen Vu[1,2], Daniel McEwen[1,2], Mark T. Edmonds[1,2], Anton Tadich[2,5] and Michael S. Fuhrer[1,2]

[1] School of Physics and Astronomy, Monash University, Melbourne, Australia
[2] ARC Centre of Excellence in Future Low Energy Electronics Technology, Monash node, Melbourne, Australia
[3] Instituto Madrileño de Estudios Avanzados en Nanociencia IMDEA Nanociencia, Madrid, Spain
[4] School of Physics and Astronomy, Queensland University of Technology, Brisbane, Australia
[5] Australian Synchrotron, Clayton, 3168, Victoria, Australia

E-mail: Iolanda.dibernardo@imdea.org and Michael.fuhrer@monash.edu

**Abstract**

The strong electron-electron interaction in transition metal dichalcogenides (TMDs) gives rise to phenomena such as strong exciton and trion binding and excitonic condensation, as well as large negative exchange and correlation contributions to the electron energies, resulting in negative electronic compressibility. Here we use angle-resolved photoemission spectroscopy to demonstrate a striking effect of negative electronic compressibility in semiconducting TMD MoS$_2$ on the charge transfer to and from a partial overlayer of monolayer semimetallic WTe$_2$. We track the changes in binding energy of the valence bands of both WTe$_2$ and MoS$_2$ as a function of surface transfer doping with donor (K) and acceptor (F4-TCNQ) species. Donor doping increases the binding energy of the MoS$_2$ valence band, as expected, while counterintuitively reducing the binding energy of the WTe$_2$ valence bands and core levels. The inverse effect is observed for acceptor doping, where a typical reduction in the MoS$_2$ binding energies is accopanied by an unexpected increase in those of WTe$_2$. The observations imply a reversal of the expected charge transfer; donor (acceptor) deposition decreases (increases) the carrier density in the WTe$_2$ adlayer. The charge transfer reversal is a direct consequence of the negative electronic compressibility of the MoS$_2$ surface layer, for which addition (subtraction) of charge leads to attraction (repulsion) of further charge from neighbouring layers. These findings highlight the importance of many-body interactions for the electrons in transition metal dichalcogenides and underscore the potential for exploring strongly correlated quantum states in two-dimensional semiconductors.

## 1. Introduction

Transition metal dichalcogenides (TMDs) have attracted considerable attention from both theoretical and application-oriented perspectives[1]. Depending on the choice of transition metal and chalcogen pair, as well as their crystallographic structure, TMDs can be metallic, semimetallic, or insulating.[2–9] As layered compounds, they feature strong in-plane covalent bonds and relatively weak out-of-plane interactions,[10,11] with electronic properties that depend on the material's thickness.[12–18] The van-der-Waals nature of these out of plane interactions allows for almost strain-free vertical heteroepitaxy of TMDs even in the presence of large lattice mismatch and different crystallographic structures[19,20]. Stacks of TMDs grown via chemical vapor deposition or molecular beam epitaxy have been proposed in applications such as field effect transistors or sensors. For these applications, understanding and controlling charge density, electronic compressibility, and charge transfer between different TMD layers is crucial.[21]

In semiconducting TMDs, the quasi-two-dimensional band structure together with the relatively large in-plane effective mass[22] leads to a small kinetic energy, while the low dielectric constant leads to a strong Coulomb interaction between electrons. As a result, the negative contributions to the electron energy due to exchange and correlation may dominate, resulting in negative electronic compressibility (NEC).[23] This is a counterintuitive phenomenon where an increase of the electron density $N$ brings a decrease of the chemical potential μ, so that the compressibility $\boldsymbol{\kappa = (1/N^2)(dN/d\mu) < 0}$.

Previous reports of NEC have primarily focused on the capacitive or electric field-screening properties of 2D electron gas-like systems like semiconductor heterostructures[24] and graphene,[25] in order to observe the effects of NEC. NEC has been observed through capacitive effects in a monolayer $MoS_2$-graphene heterostructure[22] as well as directly via angle-resolved photoemission in $WSe_2$[26] under alkali metal doping, though similar experimental observations have also been interpreted as due to a giant Stark effect.[27]

In this work, we report direct experimental evidence of NEC at the surface of bulk $MoS_2$. We use an epitaxially grown partial monolayer of $WTe_2$ on the bulk $MoS_2$ crystal as a probe: its relatively low density of states around the Fermi level allows for easy detection of any band movement induced by charge transfer to/from the substrate. By evaporating electron and hole dopants (K atoms and F4-TCNQ molecules, respectively) onto the $WTe_2/MoS_2$ surface we induce a "chemical gating" effect, aiming to reproduce the effect of field-effect doping. Shifts of the electronic band structure, both for the valence bands and core levels, are monitored via photoemission spectroscopy. We observe that incremental electron (hole) doping induces the expected progressive increase (decrease) of the binding energy of the $MoS_2$ valence band edges, while the opposite behavior is observed for the bands of the $WTe_2$ adlayer, implying a reversal of the expected direction of charge transfer. This opposite movement of the bands of the $WTe_2$ adlayer and reversal of charge transfer provide direct and unambiguous evidence of NEC at the $MoS_2$ surface. We use a simple model to elucidate how these results stem from the NEC of the topmost $MoS_2$ layer and serve as its spectroscopic fingerprint.

## 2. Methods

$WTe_2$ sub-monolayer samples were grown via molecular beam epitaxy (MBE). Commercial $MoS_2$ samples with a nominal doping of $10^{15}$ carriers $cm^{-3}$ were purchased from HQ graphene and cleaved in vacuum prior to $WTe_2$ growth. During growth, the $MoS_2$ temperature was kept at 285 °C, with an MBE chamber base pressure of $10^{-9}$ mbar. High purity tellurium (99.999%) and tungsten (99.95%) in Knudsen effusion (Kentax GmbH) and e-beam evaporation cells (FOCUS GmbH), respectively, were used as precursors. The Te Knudsen cell was kept at 245 °C during growth, yielding a rate of ~4.3 Å/min as measured via a quartz microbalance. To assess the quality of the grown film and estimate coverage, the samples were analyzed *in situ* via low-energy electron diffraction (LEED) and scanning tunnelling microscopy (STM). The samples used here for STM measurements had ~90 % $WTe_2$ ML coverage. STM measurements were performed with an electrochemically etched W tip at 77 K. Photoemission measurements were performed at the SXR beamline of the Australian Synchrotron, on samples with ~70 % $WTe_2$ ML coverage (different coverage is achieved by maintaining the same growth conditions for a different time). To avoid contaminations coming from air exposure, the samples were transferred via an ultra-high vacuum (UHV) suitcase between the two UHV systems, thus avoiding breaking vacuum. X-ray photoelectron spectroscopy (XPS) data were acquired with a photon energy of 125 eV, and Angle-resolved photoemission spectroscopy (ARPES) was performed using a He discharge lamp ($hv$ = 21.22 eV) as the photon source. All photoemission data were acquired on a toroidal analyzer (energy resolution: 150 meV). All doping and photoemission measurements were performed at 77 K, while LEED measurements were collected at room temperature. K atoms were sublimated from a commercial SAES getter; F4-TCNQ molecules purchased from Sigma-Aldrich were evaporated from a Knudsen cell. Band dispersion has been obtained by fitting 0.1 eV-wide energy distribution curves with Lorentzian curves, while the position of band maxima has been obtained by a parabolic fitting of the band dispersion. XPS and energy distribution curve (EDC) fitting was performed with the XPST tool for IgorPro, while STM images were processed with Gwyddion.

## 3. Results and discussion

### 3.1 Pristine $WTe_2/MoS_2$

Figure 1a shows the ball-stick models for the two TMDs discussed in this work, $WTe_2$ and $MoS_2$. $WTe_2$ crystallizes into a 1T' structure, with the three hexagonally packed layers of chalcogen-metal-chalcogen stacked in an ABC sequence; W atoms dimerize, resulting in a rectangular unit cell. $MoS_2$ on the other hand exhibits a 2H structure, with the chalcogen layers vertically aligned and a resultant hexagonal unit cell. Figure 1b shows a large-scale STM image of a sub-monolayer of $WTe_2$ grown via MBE on bulk $MoS_2$. The islands of $WTe_2$, characterized by dendritic edges, coalesce to form an extended interconnected network of WTe$_2$, covering about 90% of the MoS$_2$ surface. This ensures the presence of a conductive layer on the sample surface, granting the feasibility of scanning tunnelling microscopy measurements. About 20% of the WTe$_2$ layer is covered by small clusters, concentrated in the middle of the islands. We attribute such clusters to the seeding of the second layer of WTe$_2$ on top of the first one[28,29] and/or to excess precursor material which did not fully crystallize under the growth conditions used.[28]

A smaller scale STM image of the sample is reported in Figure 1d. The WTe$_2$ islands exhibit the striped pattern typical of the 1T' crystallographic structure of WTe$_2$ (see Fig. 1a). Different crystallographic grains are rotated by 120 degree angles from each other, as expected for the van der Waals epitaxy of an adlayer with a rectangular unit cell (2-fold rotational symmetry) on a substrate with a hexagonal symmetry (6-fold rotational symmetry).[28,29] A line profile across the step edge of one such island (Figure 1e) reveals an apparent step height of 850 pm, in line with the thickness of a single layer of transition metal dichalcogenides.[30] Note that this apparent height is larger than the one reported on bulk WTe$_2$[31] and MBE-grown WTe$_2$ on SrTiO3,[32] but similar to the one reported for MBE-grown WTe$_2$ on Gr/SiC[28].

The local density of states of MoS$_2$ and WTe$_2$, measured via scanning tunnelling spectroscopy at 77 K and under the same conditions for both materials, is reported in Figure 1c, for a sample with a larger WTe$_2$ coverage (about 1.5 ML). We observe the large bulk bandgap expected for semiconducting MoS$_2$, with the conduction band minimum at 0.28 eV (indicating n-doping). for WTe$_2$ a semimetallic character is observed with a density of states dip about 60 meV wide centered at approximately -70 meV. Whilst monolayer WTe$_2$ has long been established as a topological insulator,[31,33,34] it is only recently that its bulk bandgap has been demonstrated to be of excitonic origins.[35–37] An extensive discussion on the nature and size of the WTe$_2$ bulk bandgap is beyond the scope of this work.

The valence band of the WTe$_2$/MoS$_2$ system is reported in Figure 2, while that of bare MoS$_2$ in Fig. S1. We acquired angle-resolved photoemission spectra along the $\overline{\Gamma} - \overline{K}$ direction of MoS$_2$, which also cuts the BZ of one of the three WTe$_2$ orientations along the $\overline{\Gamma} - \overline{Y}$ direction. A schematic of the first Brillouin zone of our system is reported in Figure 2b, with the black hexagon corresponding to MoS$_2$ and the three colored rectangles to the three co-existing domains observed for WTe$_2$. The corresponding LEED pattern is shown in Figure 2c, with the main diffraction spots of MoS$_2$ marked by solid white arrows and the WTe$_2$ by colored dashed arrows. Prior to the growth of WTe$_2$ (Fig S1) MoS$_2$ displays a valence band maximum (VBM) at the $\overline{\Gamma}$ point, originating from the Mo-$d_{z2}$ orbitals, as expected for bulk MoS$_2$[12,38], and a local maximum at the $\overline{K}$ point originating from the Mo-$d_{x2-y2}/d_{xy}$ orbitals[12,13]. The angle-integrated spectra show a cutoff at about 1.65 eV below the Fermi level, confirming that our MoS$_2$ samples are heavily electron doped. Our experimental resolution does not, however, allow us to clearly observe the expected band splitting at the $\overline{K}$ point[27,39] nor to accurately determine the position of the VBM at $\overline{\Gamma}$. The WTe$_2$ hole-like valence band, resulting from a strong hybridization between the W-5d and Te-5p orbitals, dominates the signal closer to the Fermi level at the $\overline{\Gamma}$ point. Along the $\overline{\Gamma} - \overline{Y}$ high symmetry direction the electron-like pockets of WTe$_2$ should be visible at the Fermi level.[28,29,40] These however are best observed in ARPES experiment carried out under linearly p-polarized radiation,[29] we therefore attribute the apparent lack of such features to our experimental conditions. While the VBM of WTe$_2$ is not well resolved under our experimental conditions, the second highest valence band exhibits a much clearer dispersion; we will therefore use the position of the latter to track changes to the chemical potential for this layer in the following.

### 3.2 Electron doping

Donor doping was performed by depositing K atoms on the system at room temperature in UHV: the electron transfer from K to the WTe$_2$/MoS$_2$ system effectively results in chemical gating. The effect of doping has been estimated by monitoring the evolution of the WTe$_2$ and MoS$_2$ bands via ARPES, as seen in Figures 3a, b. We use the maxima of the EDCs to track the dispersion of the bands, as shown for the as-grown WTe$_2$/MoS$_2$ system in panel 3c. We report two exemplary peak fittings, where the position for the WTe$_2$ band (blue EDC, top panel in the inset) and MoS$_2$ band (black EDC, bottom panel in the inset) have been extracted as the centroid of the corresponding solid-filled peaks.

A comparison of panels a and b reveals that, upon doping, the MoS$_2$ bands at the $\overline{K}$ point rigidly shift towards higher binding energies. We note that, at this stage of doping, the band dispersion does not significantly vary from that of pristine MoS$_2$. We therefore neglect any considerations due to band distortion in the following. The behavior for the MoS$_2$ band, highlighted by black markers in Figure 3, is in line with the expectation that the extra electrons coming from doping occupy empty states in the conduction band causing a downward shift of the VBM[41–43] at the $\overline{K}$ point of about 200 meV as estimated by parabolic fitting.

Strikingly, we observe the opposite behavior for WTe$_2$. The VBM of the second highest WTe$_2$ band at the $\overline{\Gamma}$ point, which is marked with blue circle (squares) for undoped (K-doped) WTe$_2$, shifts about 170 meV towards lower binding energies, indicating a loss of electrons from WTe$_2$ rather than a gain, upon donor doping[44]. Subsequent K depositions only resulted in a smearing of the electronic features caused by the increased disorder in the system and did not induce any further shifts in the valence bands of either TMD, indicating that K saturation was reached already upon the first deposition. The measured valence band shifts for the K doping experiment are collected in Table 1.

**Table 1** Position of the valence band maximum for the second highest WTe$_2$ and MoS$_2$ bands before and after K doping, obtained by parabolic fitting of the band dispersion.

|  | WTe$_2$ VBM (eV) | MoS$_2$ VBM (eV) | VBM difference (eV) |
|---|---|---|---|
| *Pristine* | 0.75 | 2.00 | 1.25 |
| *K-doped* | 0.58 | 2.21 | 1.63 |

The evolution of the shallow core levels of W and Te upon potassium doping is shown in panel 3c. In both samples we identify two main sets of peaks, associated with the Te 4d (orange: Te metal, yellow: Te-W) and the W 4f core levels (blue: W-S, purple: W-Te). For the as-grown sample, the main contributions to the signal come from the WTe$_2$ component: the W 4f$_{7/2}$, located at a binding energy (BE) of 31.8 eV, and the Te 4d$_{5/2}$ at 40.6 eV, in line with previous reports[29,45]. The highest binding energy doublet, whose most intense centroid is located at 41.8 eV, is ascribed to Te atoms forming metallic clusters on top of the islands, as observed in Fig. 1b. Finally, the doublet at 33.25 eV is attributed to W in W-S bonds[46], formed between the W atoms used as precursors for WTe$_2$ and the S vacancies formed on the surface of MoS$_2$ already upon cleaving[47] or during the annealing for WTe$_2$ growth[48]. K doping (figure 3d, bottom panel) produces a shift towards lower BE of between 70 and 90 meV for all the components associated with WTe$_2$ or Te metal, as highlighted by the black vertical lines. The W-S component, on the other hand, moves 80 meV towards higher BE. This shift to lower binding energy mirrors the shift of the valence band and is opposite to the expectation for donor doping.

### 3.3 Hole doping

In a second experiment, we performed step-like deposition of the electron-acceptor molecule F4-TCNQ[49,50] on a fresh identical WTe$_2$/MoS$_2$ system, and monitored the band evolution of the WTe$_2$ and MoS$_2$ until no further change was observed in the dispersion between consecutive depositions (saturation). The results are collected in Table 2 and Figure 4.

**Table 2** Position of the valence band maximum for the second highest WTe$_2$ and MoS$_2$ bands as a function of F4-TCNQ deposition time, obtained by parabolic fitting of the band dispersion.

| F4-TCNQ Deposition time | WTe$_2$ VBM (eV) | MoS$_2$ VBM (eV) | VBM difference (eV) |
|---|---|---|---|
| 0 | 0.75 | 2.00 | 1.25 |
| 12 | 0.75 | 1.99 | 1.24 |
| 24 s | 0.84 | 1.97 | 1.13 |
| 48 s | 0.89 | 1.93 | 1.04 |

At hole-doping saturation, the MoS$_2$ bands, marked with black squares in Fig. 4a, appear to still preserve their shape while rigidly shifting about 70 meV towards lower binding energies with respect to the pristine sample (Fig. 2a), estimated by tracking the position of the VBM at the $\bar{K}$ point. While this shift is relatively small compared to our experimental resolution, the upward trend for the VBM observed in the raw data and their parabolic fitting over subsequent F4-TCNQ depositions (see Figure 4c) excludes any spurious effects associated with errors within a single measurement. Simultaneously, we observe a progressive shift of the WTe$_2$ VBM towards *higher* binding energies, as highlighted in Figure 4b. This trend is opposed to the one of MoS$_2$ and relatively clearer, as the difference in VBM position between pristine and F4-TCNQ-saturated WTe$_2$ is about 140 meV.

To further elucidate the trend in opposite directions we report, in Figure 4d, the difference between the VBM of the two species (evaluated at $\bar{\Gamma}$ for the second highest band of WTe$_2$, and at K for MoS$_2$ by means of parabolic fitting of the EDCs) as a function of F4-TCNQ deposition time. With progressive hole doping the distance between the two bands shrinks by about 200 meV, going from 1.25 eV for the pristine sample to 1.04 eV for the saturated one.

The shallow core levels for the hole-saturated sample in Figure 4e confirm the counter-intuitive trend observed for the chemical shift in the case of electron doping. All the components associated with WTe$_2$ or Te metal shift towards lower binding energies, following the trend of the WTe$_2$ valence band, while the W-S component barely moves, following the trend of the MoS$_2$ substrate.

The XPS results comparing the peaks position for the pristine sample with the electron-saturated and hole-saturated ones are collected in Table 3.

**Table 3** XPS fitting results for the W 4f and Te 5d core levels for the pristine, K saturated and F4-TCNQ saturated samples. Only the centroid of the main component for spin-orbit split doublets is reported.

|  | *Pristine (eV)* | *e-doped (eV)* | *h-doped (eV)* |
|---|---|---|---|
| W 4f$_{7/2}$ (WTe$_2$) | 31.82 | 31.74 | 31.86 |
| Te 4d$_{5/2}$ (WTe$_2$) | 40.59 | 40.52 | 40.63 |
| Te metal | 41.79 | 41.70 | 41.81 |
| W-S | 33.24 | 33.32 | 33.16 |

## 4. Discussion

The counterintuitive downward (upward) movement of the WTe$_2$ valence bands with respect to the Fermi level with positive (negative) chemical doping is due to a distinctly non-classical behavior of the substrate and requires the consideration of highly correlated electronic interactions for an accurate description. MoS$_2$, like many other TMDs, exhibits much stronger in-plane interactions than the weak van der Waals interactions out-of-plane, which separate each layer. The system can be modeled as a 2D (surface) monolayer that exhibits NEC, weakly connected to the 3D MoS$_2$ bulk and the WTe$_2$ adlayer that act classically.[22]

The evolution of these band structures under increasing electron doping is illustrated in Figure 5. In a defect-free case, there is a small contact potential between MoS$_2$ and WTe$_2$ (~130 meV,[51,52] Fig. 5a). For small donor doping, while the chemical potential lies in the gap of the MoS$_2$ at the bulk and at the surface, we expect a small positive compressibility of MoS$_2$ due to the small density of localized (trap) states in the gap. As the donor doping increases, the bands of MoS$_2$ are bent downwards at the surface until a critical level where the MoS$_2$ conduction band in the surface layer begins to populate (Fig. 5b). After this point, the electronic compressibility of the surface layer becomes negative.[22] Subsequent electron doping simultaneously populates conduction band states and lowers the chemical potential. This results in a rigid lowering of the bands relative to the surrounding WTe$_2$ adlayer and MoS$_2$ bulk bands (Fig. 5c). This draws in additional electrons to the surface layer from adjacent layers, until the built-in electrostatic potential difference cancels the drop in chemical potential. Thus, the MoS$_2$ surface layer depletes the bulk MoS$_2$, as well as the WTe$_2$ adlayer, causing a reduction in binding energy for the bands in those layers.

The fact that the reverse behavior is observed for acceptor doping suggests that the system is already in the NEC regime before deposition of any dopants. This is most likely due to the introduction of S vacancies at the MoS$_2$ surface during epitaxial growth and annealing stages which occur at elevated temperatures (285 °C). Additionally, the shallow core level spectra suggest W-S bonding which could further increase the intrinsic doping beyond the critical point. In this case, subsequent hole doping reduces the electron doping of the surface layer, increasing the chemical potential of the surface and causing transfer of electrons to neighboring layers. We attribute differences in the initial doping across samples to minor discrepancies in the sample preparation method and the age of each sample[53]. This is seen in Figure 5d, where different shaped/colored markers correspond to different samples.

Negative electronic compressibility persists over the entire range of donor and acceptor doping accessible in our experiment, corresponding to a variation in MoS$_2$ valence band binding energy of ~0.35 eV. The range of donor/acceptor density appears to be limited by saturation of the doping at high donor/acceptor concentration. While dilute concentrations of K are expected to behave as a mobile gas, large quantities of K on the surface result in the formation of metallic K clusters, where the electrons are used to form K-K bonds rather than being transferred to the MoS$_2$, resulting in saturation. As for F4-TCNQ, only the first layer contributes to charge transfer, as subsequent layers remain neutral[54].

We assume that dopants are equally distributed on the surface of WTe$_2$ and the free surface of MoS$_2$. As the WTe$_2$ monolayer is in contact with the top MoS$_2$ surface, which in the as-prepared system appears to be already in the NEC regime, charges donated to, or accepted from, the WTe$_2$ surface find their lowest energy configuration in the adjacent MoS$_2$, and the NEC behavior of MoS$_2$ attracts even more charges from the WTe$_2$. It is possible that kinetic or energetic considerations favor a preferential location for dopants on the MoS$_2$ surface compared to WTe$_2$. This would enhance the differential movement of the band structure features but cannot alone explain the reversal of charge transfer to WTe$_2$ which requires NEC in MoS$_2$.

## 5. Conclusions

We have demonstrated that negative electronic compressibility of the MoS$_2$ surface has a profound effect on the chemical doping of a partial adlayer, reversing the expected charge transfer direction, resulting in electron depletion of the adlayer on donor deposition, and electron enhancement on acceptor deposition. This striking effect provides direct and unambiguous evidence of the negative electronic compressibility of the MoS$_2$ surface. The negative compressibility effect operates over a wide range of chemical potential shifts of at least 350 meV, apparently limited only by the saturation of charge transfer from the donor/acceptor layers. The observation of a large negative electronic compressibility in a MoS$_2$/WTe$_2$ heterostructure may inform the design of novel electronic devices with enhanced, or non-linear capacitances.

## Acknowledgements

I.D.B. acknowledges support from MSCA Program (101063547-GAP-101063547). M.T.E. acknowledges funding support from ARC Future Fellowship (FT2201000290). J.B. acknowledges support from AINSE Ltd. Postgraduate Research Award (PGRA). All authors acknowledge funding from the FLEET Centre of Excellence, ARC Grant No. CE170100039. IMDEA Nanociencia and IFIMAC acknowledge financial support from the Spanish Ministry of Science and Innovation 'Severo Ochoa' (Grant CEX2020-001039-S) and 'María de Maeztu' (Grant CEX2018-000805-M) Programme for Centers of Excellence in R&D, respectively. Part of this research was undertaken on the soft x-ray spectroscopy beamline at the Australian Synchrotron, ANSTO.

**References**

1. Wang, Q. H., Kalantar-Zadeh, K., Kis, A., Coleman, J. N. & Strano, M. S. Electronics and optoelectronics of two-dimensional transition metal dichalcogenides. *Nat. Nanotechnol.* **7**, 699–712 (2012).
2. Wilson, J. A. & Yoffe, A. D. The transition metal dichalcogenides discussion and interpretation of the observed optical, electrical and structural properties. *Adv. Phys.* **18**, 193–335 (1969).
3. Rajan, A., Underwood, K., Mazzola, F. & King, P. D. C. C. Morphology control of epitaxial monolayer transition metal dichalcogenides. *Phys. Rev. Mater.* **4**, 014003 (2020).
4. Voiry, D., Mohite, A. & Chhowalla, M. Phase engineering of transition metal dichalcogenides. *Chem. Soc. Rev.* **44**, 2702–2712 (2015).
5. Komsa, H. & Krasheninnikov, A. V. Engineering the Electronic Properties of Two-Dimensional Transition Metal Dichalcogenides by Introducing Mirror Twin Boundaries. *Adv. Electron. Mater.* **3**, 1–10 (2017).
6. Yokoya, T. *et al.* Fermi Surface Sheet-Dependent Superconductivity in 2 H -NbSe 2. *Science (80-. ).* **294**, 2518–2520 (2001).
7. Sipos, B. *et al.* From Mott state to superconductivity in-1T-TaS2. *Nat. Mater.* **7**, 960–965 (2008).
8. Qian, X., Liu, J., Fu, L. & Li, J. Quantum spin Hall effect in two-dimensional transition metal dichalcogenides. *Science (80-. ).* **346**, 1344–1347 (2014).
9. Manzeli, S., Ovchinnikov, D., Pasquier, D., Yazyev, O. V. & Kis, A. 2D transition metal dichalcogenides. *Nat. Rev. Mater.* **2**, (2017).
10. Frindt, R. F. & Yoffe, A. D. Physical properties of layer structures : optical properties and photoconductivity of thin crystals of molybdenum disulphide. *Proc. R. Soc. London. Ser. A. Math. Phys. Sci.* **273**, 69–83 (1963).
11. Novoselov, K. S. *et al.* Two-dimensional atomic crystals. *Proc. Natl. Acad. Sci. U. S. A.* **102**, 10451–10453 (2005).
12. Jin, W. *et al.* Direct measurement of the thickness-dependent electronic band structure of MoS2 using angle-resolved photoemission spectroscopy. *Phys. Rev. Lett.* **111**, 106801 (2013).
13. Cappelluti, E., Roldán, R., Silva-Guillén, J. A., Ordejón, P. & Guinea, F. Tight-binding model and direct-gap/indirect-gap transition in single-layer and multilayer MoS2. *Phys. Rev. B* **88**, 075409 (2013).
14. Fei, Z. *et al.* Edge conduction in monolayer WTe2. *Nat. Phys.* **13**, 677–682 (2017).
15. Villaos, R. A. B. *et al.* Thickness dependent electronic properties of Pt dichalcogenides. *npj 2D Mater. Appl.* **3**, 1–8 (2019).
16. Pang, C. S., Wu, P., Appenzeller, J. & Chen, Z. Thickness-Dependent Study of High-Performance WS2-FETs with Ultrascaled Channel Lengths. *IEEE Trans. Electron Devices* **68**, 2123–2129 (2021).
17. Zhang, Y. *et al.* Direct observation of the transition from indirect to direct bandgap in atomically thin epitaxial MoSe2. *Nat. Nanotechnol.* **9**, 111–115 (2014).
18. Chen, R.-S., Tang, C.-C., Shen, W.-C. & Huang, Y.-S. Thickness-dependent electrical conductivities and ohmic contacts in transition metal dichalcogenides multilayers. *Nanotechnology* **25**, 415706 (2014).
19. Gong, Y. *et al.* Vertical and in-plane heterostructures from WS2/MoS2 monolayers. *Nat. Mater.* **13**, 1135–1142 (2014).
20. Liang, J. *et al.* Controlled Growth of Two-Dimensional Heterostructures: In-Plane Epitaxy or Vertical Stack. *Accounts Mater. Res.* **3**, 999–1010 (2022).
21. Park, J. H. *et al.* Defect passivation of transition metal dichalcogenides via a charge transfer van der Waals interface. *Sci. Adv.* **3**, e1701661 (2017).
22. Larentis, S. *et al.* Band offset and negative compressibility in graphene-MoS2 heterostructures. *Nano Lett.* **14**, 2039–2045 (2014).
23. Eisenstein, J. P., Pfeiffer, L. N. & West, K. W. Negative compressibility of interacting two-dimensional electron and quasiparticle gases. *Phys. Rev. Lett.* **68**, 674–677 (1992).
24. Lu Li, C. Richter, S. Paetel, T. Kopp, J. Mannhart, R. C. A. Very Large Capacitance Enhancement in a Two-Dimensional Electron System. *Science (80-. ).* **332**, 825 (2011).
25. Yu, G. L. *et al.* Interaction phenomena in graphene seen through quantum capacitance. *Proc. Natl. Acad. Sci.* **110**, 3282–3286 (2013).
26. Riley, J. M. *et al.* Negative electronic compressibility and tunable spin splitting in WSe 2. *Nat. Nanotechnol.* **10**, 1043–1047 (2015).
27. Kang, M. *et al.* Universal Mechanism of Band-Gap Engineering in Transition-Metal Dichalcogenides. *Nano Lett.* **17**, 1610–1615 (2017).
28. Llopez, A. *et al.* Van der Waals Epitaxy of Weyl-Semimetal Td-WTe2. *ACS Appl. Mater. Interfaces* (2024) doi:10.1021/acsami.4c00676.
29. Tang, S. *et al.* Quantum spin Hall state in monolayer 1T'-WTe 2 - Supplimentary Information. *Nat. Phys.* **13**, 683–687 (2017).
30. Di Bernardo, I. *et al.* Metastable Polymorphic Phases in Monolayer TaTe2. *Small* **19**, (2023).
31. Peng, L. *et al.* Observation of topological states residing at step edges of WTe2. *Nat. Commun.* **8**, 1–7 (2017).
32. Li, H. *et al.* Molecular beam epitaxy growth and strain-induced bandgap of monolayer 1T′-WTe2 on SrTiO3(001). *Appl. Phys. Lett.* **117**, (2020).
33. Shi, Y. *et al.* Imaging quantum spin Hall edges in monolayer WTe 2. *Sci. Adv.* **5**, 1–7 (2019).
34. Jia, Z. Y. *et al.* Direct visualization of a two-dimensional topological insulator in the single-layer 1T'-WT e2 - Supplementary materials\. *Phys. Rev. B* **1**, 100 (2017).
35. Sun, B. *et al.* Evidence for equilibrium exciton condensation in monolayer WTe2. *Nat. Phys.* (2021) doi:10.1038/s41567-021-01427-5.
36. Jia, Y. *et al.* Evidence for a monolayer excitonic insulator. *Nat. Phys.* **18**, 87–93 (2022).
37. Que, Y. *et al.* A Gate-Tunable Ambipolar Quantum Phase Transition in a Topological Excitonic Insulator. *Adv. Mater.* **36**, (2024).
38. Mahatha, S. K., Patel, K. D. & Menon, K. S. R. Electronic structure investigation of MoS2 and MoSe2 using angle-resolved photoemission spectroscopy and ab initio band structure studies. *J. Phys. Condens. Matter* **24**, 475504 (2012).
39. Kim, B. S., Rhim, J.-W., Kim, B., Kim, C. & Park, S. R. Determination of the band parameters of bulk 2H-MX2 (M = Mo, W; X = S, Se) by angle-resolved photoemission spectroscopy. *Sci. Rep.* **6**, 36389 (2016).
40. Cucchi, I. *et al.* Microfocus Laser-Angle-Resolved Photoemission on Encapsulated Mono-, Bi-, and Few-

Layer 1T′-WTe 2. *Nano Lett.* **19**, 554–560 (2019).
41. Miwa, J. A. *et al.* Electronic Structure of Epitaxial Single-Layer MoS 2. **046802**, 1–5 (2015).
42. Eads, C. N. *et al.* Ultrafast Carrier Dynamics in Two-Dimensional Electron Gas-like K-Doped MoS2. *J. Phys. Chem. C* **124**, 19187–19195 (2020).
43. Komesu, T. *et al.* Occupied and unoccupied electronic structure of Na doped MoS2(0001). *Appl. Phys. Lett.* **105**, 241602 (2014).
44. Xia, Y. *et al.* Hole doping in epitaxial MoSe 2 monolayer by nitrogen plasma treatment Hole doping in epitaxial MoSe 2 monolayer by nitrogen plasma treatment. (2018).
45. Walsh, L. A. *et al.* WTe2 thin films grown by beam-interrupted molecular beam epitaxy. *2D Mater.* **4**, 0–6 (2017).
46. Jaegermann, W., Ohuchi, F. S. & Parkinson, B. A. Interaction of Cu, Ag and Au with van der Waals faces of WS, and SnS2. *Surf. Sci.* **201**, 211–227 (1988).
47. Di Bernardo, I. *et al.* Defects, band bending and ionization rings in MoS2. *J. Phys. Condens. Matter* **34**, (2022).
48. Lu, W., Birmingham, B. & Zhang, Z. Defect engineering on MoS2 surface with argon ion bombardments and thermal annealing. *Appl. Surf. Sci.* **532**, 147461 (2020).
49. Edmonds, M. T., Hellerstedt, J., O'Donnell, K. M., Tadich, A. & Fuhrer, M. S. Molecular Doping the Topological Dirac Semimetal Na 3 Bi across the Charge Neutrality Point with F4-TCNQ. *ACS Appl. Mater. Interfaces* **8**, 16412–16418 (2016).
50. Pinto, H., Jones, R., Goss, J. P. & Briddon, P. R. p-type doping of graphene with F4-TCNQ. *J. Phys. Condens. Matter* **21**, 402001 (2009).
51. Schlaf, R., Lang, O., Pettenkofer, C. & Jaegermann, W. Band lineup of layered semiconductor heterointerfaces prepared by van der Waals epitaxy: Charge transfer correction term for the electron affinity rule. *J. Appl. Phys.* **85**, 2732–2753 (1999).
52. Torun, E., Sahin, H., Cahangirov, S., Rubio, A. & Peeters, F. M. Anisotropic electronic, mechanical, and optical properties of monolayer WTe2. *J. Appl. Phys.* **119**, 0–7 (2016).
53. Park, Y. *et al.* Unveiling the origin of n-type doping of natural MoS2: carbon. *npj 2D Mater. Appl.* **7**, 60 (2023).
54. Chen, W., Chen, S., Qi, D. C., Gao, X. Y. & Wee, A. T. S. Surface Transfer p-Type Doping of Epitaxial Graphene. *J. Am. Chem. Soc.* **129**, 10418–10422 (2007).
55. Jain, A. *et al.* Commentary: The Materials Project: A materials genome approach to accelerating materials innovation. *APL Mater.* **1**, (2013).

# FIGURES

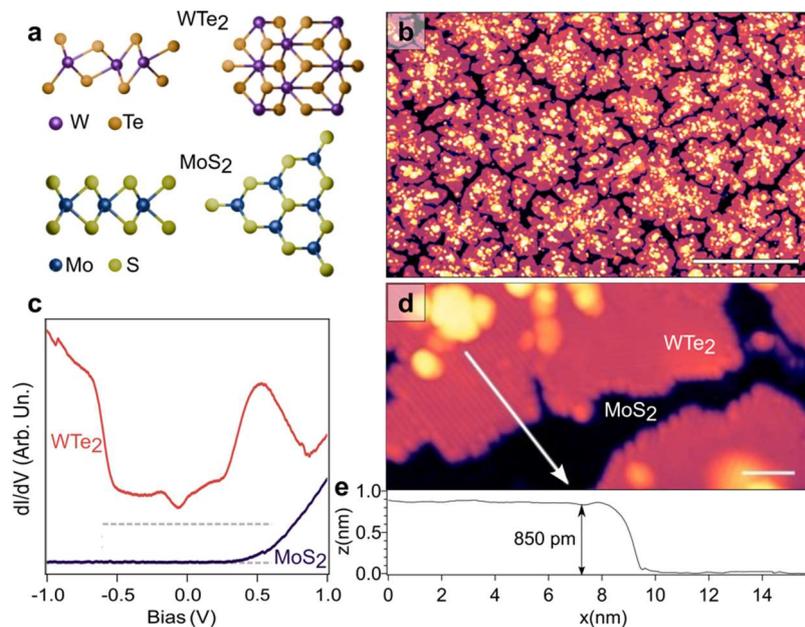

**Figure 1** *Real space characterization of WTe$_2$/MoS$_2$. (a) Ball-stick atomic model for the side (left) and top view (right) of WTe$_2$ (top row) and MoS$_2$ (bottom row). (b) (400x200) nm STM topography of WTe$_2$/MoS$_2$; I = 20 pA, V = 1 V, scalebar = 50 nm. (c) dI/dV curves acquired on a sample with 1.5 WTe$_2$ ML coverage and on its MoS$_2$ substrate; V$_{mod}$ = 10 mV, f$_{mod}$ = 828 Hz, I$_{set}$ = 20 pA. (d) (40x20) nm STM topography close up on an island of WTe$_2$, I = 15 pA, V = 1 V; (e) line profile across the white arrow in top panel. Ball-stick atomic models reproduced from The Materials Project database.*[55]

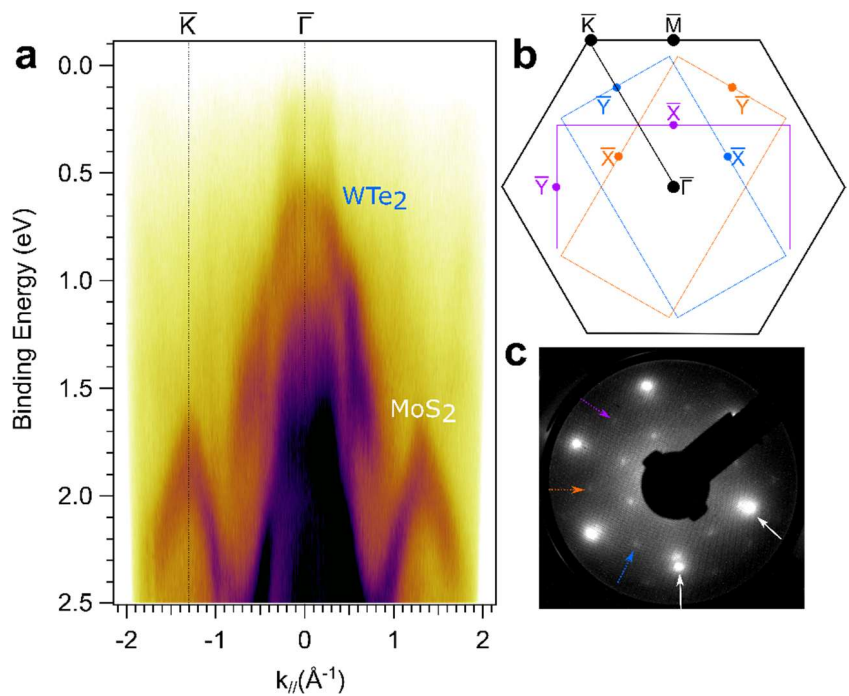

**Figure 2** *Reciprocal space characterization of WTe$_2$/MoS$_2$. (a) Valence band dispersion of WTe$_2$/MoS$_2$ along the Γ-K direction of MoS$_2$; the high symmetry points of MoS$_2$ are marked by vertical dashed lines. (b) Model of the reciprocal unit cell of the WTe$_2$/MoS$_2$ system. (c) LEED pattern corresponding to the sample in (a); beam energy: 80 eV.*

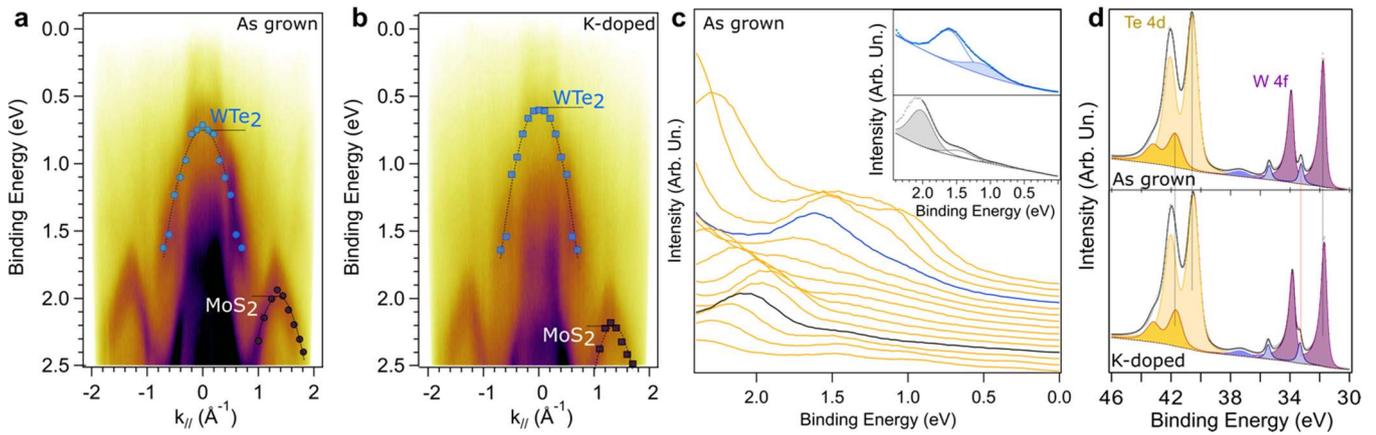

**Figure 3** *Effect of electron doping. (a) Valence band dispersion of WTe$_2$/MoS$_2$ of along the Γ-K direction of MoS$_2$. Here and in (b) the dispersion of the WTe$_2$ (MoS$_2$) band is marked by blue (black) markers, with parabolic fittings superimposed as black dotted lines. (b) Valence band dispersion of WTe$_2$/MoS$_2$ along the Γ-K direction of MoS$_2$ at k-doping saturation. (c) EDC dispersion for panel (a) using a 0.1 Å$^{-1}$ integration step size. Top inset: peak fitting corresponding to the blue EDC profile, where the solid blue peak corresponds to the WTe$_2$ band. Bottom inset: peak fitting for the grey profile, where the solid grey peak corresponds to the MoS$_2$ band. (d) Te 4d and W 4f core level for the undoped (top) and saturation K-doped (bottom) system. Peak assignment: Te metal (orange), Te-W (yellow), W-S (blue), W-Te (purple).*

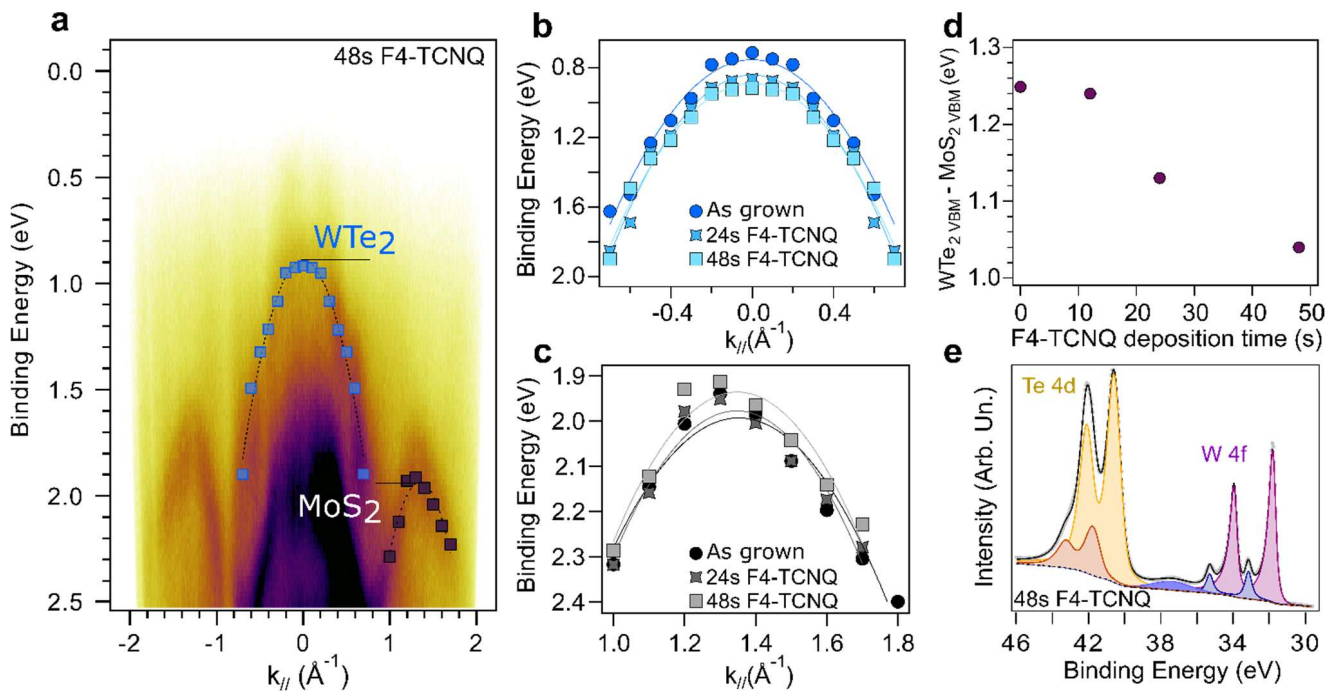

**Figure 4** *Effect of hole doping. (a) Valence band dispersion of WTe$_2$/MoS$_2$ along the Γ-K direction of MoS$_2$ at F4-TCNQ doping saturation. The dispersion of the WTe$_2$ (MoS$_2$) band is marked by blue (black) markers, with parabolic fittings superimposed as black dotted lines. (b) and (c); evolution of the WTe$_2$ and MoS$_2$ bands, respectively, as a function of F4-TCNQ doping. (d) evolution of the distance between the VBM of WTe$_2$ and the VBM of MoS$_2$ as a function of F4-TCNQ doping, demonstrating that hole-doping causes the bands to move in opposite directions. (e) Te 4d and W 4f core levels at hole-doping saturation. Peak assignment: Te metal (orange), Te-W (yellow), W-S (blue), W-Te (purple).*

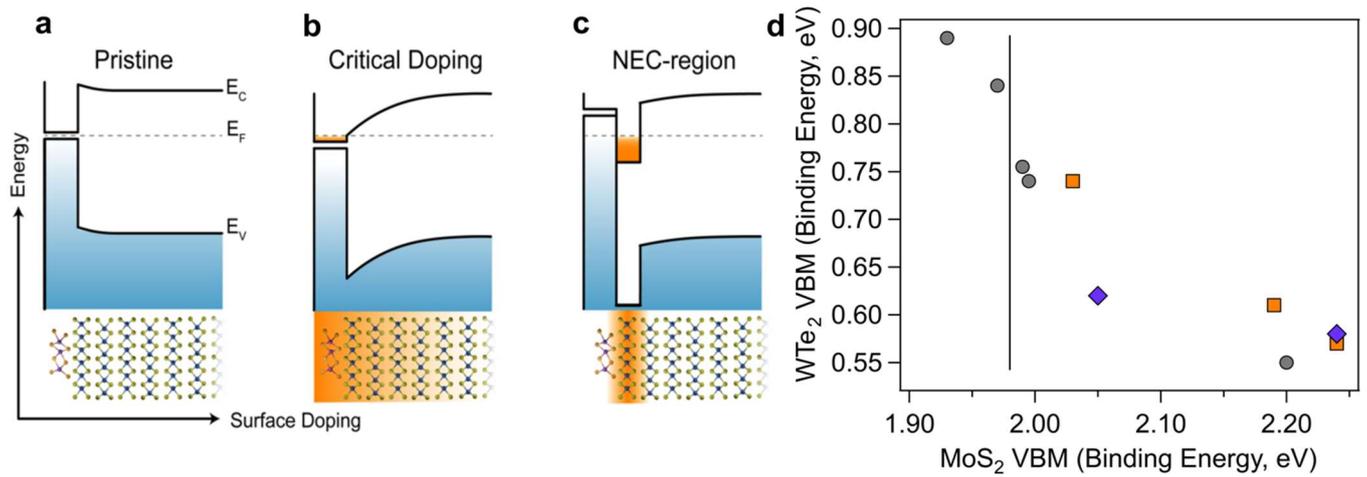

**Figure 5** *Negative electronic compressibility on the topmost layer of MoS$_2$. (a)-(c) schematic of band alignment for a pristine WTe$_2$/MoS$_2$ system as a function of intrinsic surface doping. A cross-section of the WTe$_2$/MoS$_2$ system is aligned below, with a coloured gradient indicating the approximate distribution of charge. E$_C$ = conduction band, E$_V$ = valence band, E$_F$ = Fermi level. (d) position of the second topmost valence band maximum of WTe$_2$ as a function of the valence band maximum of the underlying MoS$_2$ for different levels of electrons and hole doping. Different markers represent different doping experiments (i.e., different WTe$_2$/MoS$_2$ crystals doped with either electrons or holes) the vertical line marks the position of the VBM for undoped MoS$_2$.*

# Supporting Information - Reversal of charge transfer doping on the negative electronic compressibility surface of MoS₂

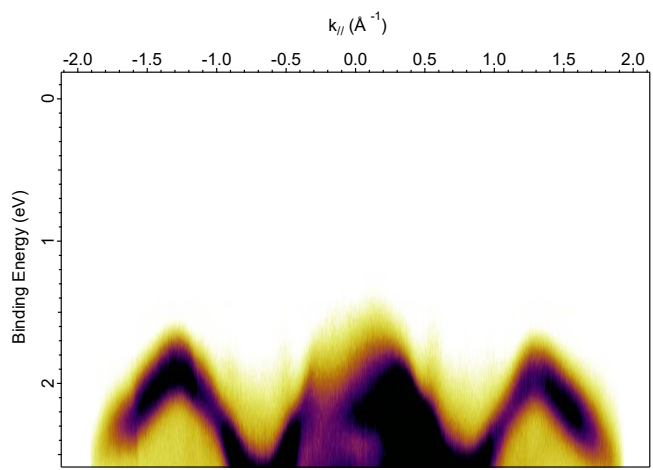

**Figure S1:** *Band structure of freshly cleaved MoS2 along the K–Γ–K direction. While the VBM at the $\bar{K}$ points are well defined, the VBM at the $\bar{\Gamma}$ point is not defined as well due to the experimental conditions (polarization, photon energy). Photon energy: 21.22 eV.*

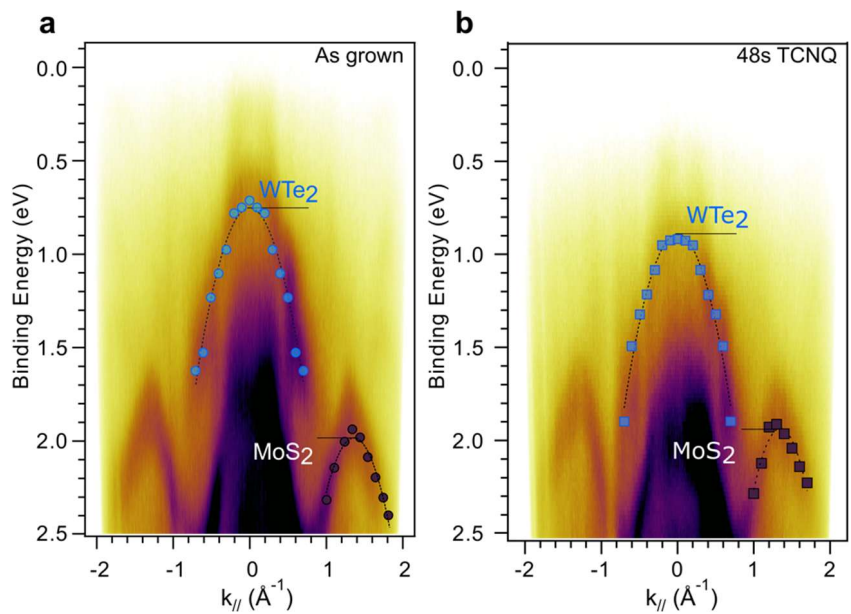

**Figure S2:** *Effect of hole doping. (a) Valence band dispersion of WTe2/MoS2 of along the $\bar{\Gamma} - \bar{K}$ direction of MoS2. Here and in (b) the dispersion of the WTe2 (MoS2) band is marked by blue (black) markers, with parabolic fittings superimposed as black dotted lines. (b) Valence band dispersion WTe2/MoS2 of along the $\bar{\Gamma} - \bar{K}$ direction of MoS2 at 48 s of F4-TCNQ doping. Photon energy: 21.22 eV.*

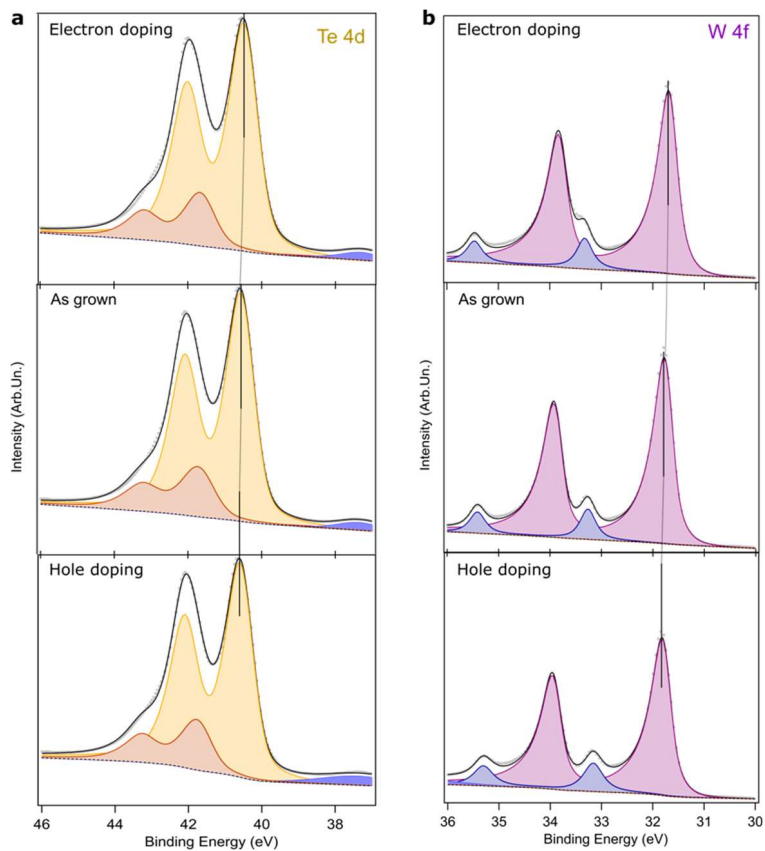

**Figure S3:** *(a) Te 4d and (b) W 4f core levels comparing the relative binding energy shifts between (top) electron doping saturation (middle) as grown, and (bottom) hole-doping saturation conditions. Grey dotted lines are guides to the eye illustrating the shift in the peak centroid (black vertical lines) with chemical doping. Peak assignment: Te metal (orange), Te-W (yellow), W-S (blue), W-Te (purple). Photon energy: 125 eV.*